\documentclass[aps,prc,preprint,groupedaddress,showpacs]{revtex4}
\usepackage{amsmath}
\usepackage{amssymb}
\usepackage{bm}
\usepackage{epsfig, graphicx, euscript}
\newcommand{\be}{\begin{equation}}
\newcommand{\ee}{\end{equation}}
\newcommand{\bea}{\begin{eqnarray}}
\newcommand{\eea}{\end{eqnarray}}

\newcommand{\mbts}[1]{_{\mbox{\tiny #1}}}
\newcommand{\mbsu}[1]{\mbox{\scriptsize #1}}

\begin{document}
\title{
ELIMINATION OF $\mathbf{0^+}$ SPURIOUS STATES IN THE
QUASIPARTICLE TIME BLOCKING APPROXIMATION}

\author{V.~I.~Tselyaev}
\affiliation{Nuclear Physics Department, V. A. Fock Institute of Physics,
St. Petersburg State University, 198504 St. Petersburg, Russia}

\date{\today}

\begin{abstract}
The quasiparticle time blocking approximation (QTBA)
is considered as a model for the description of excitations
in open-shell nuclei. The QTBA is an extension of
the quasiparticle random phase approximation
that includes quasiparticle-phonon coupling.
In the present version of the QTBA,
the pairing correlations are included within the framework
of the BCS approximation. Thus, in this model, the $0^+$ spurious
states appear, which are caused by the breaking of
the symmetry related to the particle-number conservation.
In this work, the method is described which solves
the problem of the $0^+$ spurious states in the QTBA
with the help of the projection technique.
The method is illustrated
by calculations of $0^+$ excitations in $^{120}$Sn nucleus.
\end{abstract}


\email{tselyaev@nuclpc1.phys.spbu.ru}

\maketitle

\section{Introduction}

Description of nuclear excitations within the models
based on the mean-field approach faces the known difficulty
of appearing the so-called spurious or ghost states
(see, e.~g., \cite{RS80}).
The spurious state arises because of the breaking of some symmetry.
In particular, the breaking of the translation symmetry leads
to the appearance of spurious $1^-$ states.
In the fully self-consistent theory, for instance,
in the self-consistent random phase approximation (RPA)
or in the self-consistent quasiparticle RPA (QRPA),
the broken symmetry is restored and the spurious states disappear
\cite{RS80,T61,KS82}.
The self-consistency means here a fulfillment of some constraints
imposed on the mean-field operator and the amplitude of the
residual interaction.
However, it is difficult to achieve the full self-consistency
in the models which go beyond the (Q)RPA framework
by taking into account additional correlations.
So, the problem of the spurious states becomes relevant again,
even if it was solved on the (Q)RPA level.
In this work, the quasiparticle time blocking approximation
(QTBA, see Refs. \cite{QTBA1,QTBA2}) is considered.
The QTBA is a model intended for the description of excitations
in open-shell nuclei.
The model is formulated in terms of the Green function method.
Within the QTBA, the single-particle continuum,
the pairing correlations, and the quasiparticle-phonon coupling
(QPC) are included. In this sense, the QTBA is an extension
of the QRPA. Note that all three components of the model
mentioned above are necessary to describe the fragmentation
and the width of the states in open-shell nuclei.
In the present version of the QTBA,
the pairing correlations are included within the framework
of the Bardeen-Cooper-Schrieffer (BCS) approximation.
Thus, in this model,
the $0^+$ spurious states appear, which are caused
by the breaking of the symmetry related to the
particle-number conservation. In the QRPA,
the $0^+$ spurious state is eliminated by taking into account
the so-called dynamical pairing effects
(particle-particle channel contributions, see Ref.~\cite{M67}).
However, in the QTBA, this is not sufficient because
the spurious state is fragmented due to its coupling
to the ``two-quasiparticle$\otimes$phonon''
(2q$\otimes$phonon) configurations.
In this work, the method is described which solves
the problem of the $0^+$ spurious states in the QTBA
with the help of a combination of the so-called
subtraction procedure (see \cite{QTBA1,QTBA2})
and the projection technique described below.
The scheme proposed is illustrated by the calculations
of $0^+$ excitations in $^{120}$Sn nucleus.

\section{System of the QTBA equations and its modifications}

The basic equation,
which has to be solved within the Green function method
to calculate the physical observables related to the nuclear excitations,
is the equation for the effective (renormalized) response function
$R^{\mbsu{eff}}(\omega)$.
It has the same form both in the (Q)RPA and in the QTBA.
In the symbolic notations it reads
(in what follows we will use notations and definitions
of Ref.~\cite{QTBA2}):
\be
R^{\mbsu{eff}}(\omega) = A(\omega) - A(\omega)\,{\cal F}\,
R^{\mbsu{eff}}(\omega)
\label{bse1}
\ee
where $A(\omega)$ is a correlated propagator
and $\cal F$ is an amplitude of the effective residual interaction.
In the case of the QRPA, $A(\omega)$ reduces to the uncorrelated
2q propagator $\tilde{A}(\omega)$.
In the general case including pairing correlations,
the amplitude $\cal F$ can be represented as a sum of two terms
\be
{\cal F} = {\cal F}^{(\mbsu{ph})} + {\cal F}^{(\mbsu{pp})}
\label{fsum}
\ee
where the amplitude ${\cal F}^{(\mbsu{ph})}$ represents interaction
in the particle-hole (ph) channel and ${\cal F}^{(\mbsu{pp})}$
includes contributions of the interaction
both in the particle-particle (pp) and in the hole-hole (hh) channels
(in the following for brevity we will use the unified term
pp channel implying also the hh-channel contributions).
The response function $R^{\mbsu{eff}}(\omega)$ enables one to
calculate the strength function $S(E)$ which determines
the distribution of the transition strength caused by an
external field $V^0$. These quantities are related by the following
formulas:
\be
S(E) = -\frac{1}{\pi}\,\mbox{Im}\,\Pi (E + i \Delta)\,,
\label{defsf}
\ee
\be
\Pi (\omega) = -\frac{1}{2}\,\mbox{Tr}\,\bigl(
(eV^{\,0})^{\dag}\,R^{\,\mbsu{eff}}(\omega)\,(eV^{\,0})\bigr)\,,
\label{defpol}
\ee
where $\Pi (\omega)$ is the nuclear polarizability,
$E$ is an excitation energy,
$\Delta$ is a smearing parameter, and
$e$ is an effective charge operator.

One of the important questions arising in the QRPA and QTBA calculations
is the question of completeness of the configuration space.
The size of the basis in this space has an impact practically on all
the calculated quantities. In particular, configurations with a particle
in the continuum are responsible for the formation of the escape widths
of the resonances.
The well-known method to include these configurations on the RPA level
is the use of the coordinate representation within the Green function
formalism (see Ref.~\cite{SB75}). This method is used in the present
approach as described in Ref.~\cite{QTBA2}.
However, incorporation of the pp-channel contributions in
the coordinate representation leads to considerable numerical
difficulties. At the same time, the pp-channel contributions
(dynamical pairing effects) are very important in the
calculations of $0^+$ excitations in the open-shell nuclei,
primarily because of the problem of the $0^+$ spurious state.
For this reason, in Ref.~\cite{gmrQTBA} a combined method
was developed, which is a modification of the so-called
$(r,\lambda)$ representation proposed in Ref.~\cite{PS88}
for the QRPA problem. Within this method only the ph channel
is treated in the coordinate space; the dynamical pairing effects
are included in the discrete basis representation.

Consider the general case of the QTBA.
By taking into account the decomposition (\ref{fsum}) one can rewrite
Eq.~(\ref{bse1}) in the form
\be
R^{\mbsu{eff}}(\omega) = A^{(\mbsu{res+pp})}(\omega)
- A^{(\mbsu{res+pp})}(\omega)\,{\cal F}^{(\mbsu{ph})}\,
R^{\mbsu{eff}}(\omega)
\label{bse2}
\ee
where propagator $A^{(\mbsu{res+pp})}(\omega)$ is a solution
of the equation
\be
A^{(\mbsu{res+pp})}(\omega) = A(\omega)
- A(\omega)\,{\cal F}^{(\mbsu{pp})}\,
A^{(\mbsu{res+pp})}(\omega)\,.
\label{app1}
\ee
In the present work the version of the QTBA is used in which
the ground state correlations caused by the QPC are neglected.
In this case the correlated propagator $A(\omega)$ is defined
by the equation
\be
A(\omega) = \tilde{A}(\omega)
- \tilde{A}(\omega)\,\bar{\Phi}(\omega)\,A(\omega)
\label{cprp}
\ee
where $\tilde{A}(\omega)$ is the uncorrelated QRPA  propagator,
\be
\bar{\Phi}(\omega) =
\Phi^{(\mbsu{res})}(\omega) - \Phi^{(\mbsu{res})}(0)\,,
\label{bphi}
\ee
and $\Phi^{(\mbsu{res})}(\omega)$ is a resonant part of the
interaction amplitude responsible for the QPC in the QTBA
(see Refs.~\cite{QTBA2,QTBA1} for details).
Combining Eqs. (\ref{app1}) and (\ref{cprp}) leads to the
new equation for $A^{(\mbsu{res+pp})}(\omega)$:
\be
A^{(\mbsu{res+pp})}(\omega) = \tilde{A}(\omega) - \tilde{A}(\omega)
\bigl[\bar{\Phi}(\omega) + {\cal F}^{(\mbsu{pp})}\bigr]
A^{(\mbsu{res+pp})}(\omega)\,.
\label{app2}
\ee

As a result we find that the pp-channel contributions can be included
by modification of the equation for the correlated propagator, i.~e.
by replacing Eq.~(\ref{cprp}) by Eq.~(\ref{app2}). The modification
is reduced to the additional term ${\cal F}^{(\mbsu{pp})}$ added to
the amplitude $\bar{\Phi}(\omega)$.

Notice, however, that in practice Eq.~(\ref{bse2}) for
$R^{\mbsu{eff}}(\omega)$ is solved in the coordinate representation
(to take into account the single-particle continuum),
whereas Eq.~(\ref{app2})
is solved in the restricted discrete basis representation.
This fact greatly simplifies the problem as compared with the initial
Eq.~(\ref{bse1}) in which both the ph-channel contribution and
the pp-channel one are included in the coordinate representation.
At the same time, the use of the restricted discrete basis
representation for the pp channel is fully consistent with the
BCS approximation in which the gap equation is solved in the same
restricted basis. In more detail, this modification of the QTBA
equations is described in Ref.~\cite{gmrQTBA}.

\section{The method of eliminating the {\boldmath $\,0^+$}
spurious state}

The general scheme described above ensures that the energy of the
$0^+$ spurious (ghost) state is equal to zero both in the QRPA
and in the QTBA. Indeed,
it is not difficult to show that Eq.~(C1) of Ref.~\cite{QTBA2}
for the transition amplitudes of the spherically symmetric nucleus
in the QRPA has a non-zero solution at the energy $\omega_q = 0$
and at the total angular momentum and the parity $J^{\pi}_q = 0^+$
if the gap equation (A25) is fulfilled.
The explicit form of this solution is
\be
\rho^{0^+(\mbsu{ghost})}_{(12)\eta} =
C\,\delta^{\vphantom{(+)}}_{(12)}\,\rho^{\vphantom{(+)}}_{(1)\eta}\,,
\label{rgh1}
\ee
where $C$ is an arbitrary constant,
\be
\rho^{\vphantom{(+)}}_{(1)\eta} =
\eta\,\rho^{\vphantom{(+)}}_{(1)}\,,\qquad
\rho^{\vphantom{(+)}}_{(1)} = \sqrt{2j^{\vphantom{(J)}}_1+1}\,
u^{\vphantom{(J)}}_{(1)}v^{\vphantom{(J)}}_{(1)}\,,\qquad
\eta = \pm 1\,.
\label{rgh2}
\ee
Here and in the following it is supposed that the equations of the
model are written in the representation of
the single-quasiparticle basis functions in the doubled space
$\tilde{\psi}^{\vphantom{*}}_1$ which
(according to the notations of Refs.~\cite{QTBA1,QTBA2})
are labelled by the composite indices
$1 = \{[1],m^{\vphantom{*}}_1\}$
where
$[1] = \{(1),\eta^{\vphantom{*}}_1\}$,
$(1)=\{\tau^{\vphantom{*}}_1,n^{\vphantom{*}}_1,l^{\vphantom{*}}_1,
j^{\vphantom{*}}_1\}$, and $\eta^{\vphantom{*}}_1 = \pm 1$
is the sign of the quasiparticle energy
$E^{\vphantom{*}}_1 = \eta^{\vphantom{*}}_1 E^{\vphantom{*}}_{(1)}$.
That is, the symbol ``$(1)$'' stands for the set of the single-particle
quantum numbers excepting the projection of the total angular momentum
$m^{\vphantom{*}}_1$, $v^2_{(1)}$ is the occupation probability,
and $u^{\vphantom{*}}_{(1)}=\sqrt{1-v^2_{(1)}}$.

The existence of the solution (\ref{rgh1})
means that the response function $R^{\mbsu{eff}}(\omega)$
as a solution of Eq.~(\ref{bse2}) in the QRPA has a pole at $\omega = 0$.
The same is true for the QTBA since at $\omega = 0$
the QTBA equation (\ref{bse2}) coincides with the QRPA one owing to
the subtraction procedure determined by Eq.~(\ref{bphi}).

However, there still remains the following problem:
in the QTBA the ghost state can be fragmented due to its coupling
to the 2q$\otimes$phonon configurations,
despite the energy of the dominant ghost state is equal to zero.
It can lead to the spurious states at low energies distorting respective
strength functions. In particular, these fragmented spurious states
will produce non-zero response to the particle-number operator
which has to be exactly equal to zero in a correct theory
(as, for instance, in the QRPA including pp channel that was proved
by Migdal, see \cite{M67}).

To solve this problem, the following method is proposed.
Let us recast Eq.~(\ref{app2}) in the form
\be
A^{(\mbsu{res+pp})}(\omega) = \tilde{A}^{(\mbsu{pp})}(\omega) -
\tilde{A}^{(\mbsu{pp})}(\omega)\,\bar{\Phi}(\omega)\,
A^{(\mbsu{res+pp})}(\omega)\,,
\label{app3}
\ee
where propagator $\tilde{A}^{(\mbsu{pp})}(\omega)$ is a solution
of the equation:
\be
\tilde{A}^{(\mbsu{pp})}(\omega) = \tilde{A}(\omega) -
\tilde{A}(\omega)\,{\cal F}^{(\mbsu{pp})}\,
\tilde{A}^{(\mbsu{pp})}(\omega)\,.
\label{appt1}
\ee
Consider the case $J^{\pi}_q = 0^+$ assuming diagonal approximation
for the amplitude ${\cal F}^{(\mbsu{pp})}$ (which is consistent
with the gap equation in the BCS approximation, see Appendix~C
of Ref.~\cite{gmrQTBA}).
In this case one can keep only diagonal
parts of the matrix functions in Eq.~(\ref{appt1}).
In the explicit form we have
\be
\tilde{A}^{(\mbsu{pp})}_{(1)\eta,\,(2)\eta'}(\omega) =
\tilde{A}^{\vphantom{(\mbsu{pp})}}_{(1)\eta,\,(2)\eta'}(\omega)
- \sum_{(34)\eta''\eta'''}
\tilde{A}^{\vphantom{(\mbsu{pp})}}_{(1)\eta,\,(3)\eta''}(\omega)\,
{\cal F}^{J=0\,(\mbsu{pp})}_{(33)\eta'',\,(44)\eta'''}\,
\tilde{A}^{(\mbsu{pp})}_{(4)\eta''',\,(2)\eta'}(\omega)
\label{appt2}
\ee
where
$\tilde{A}^{(\mbsu{pp})}_{(1)\eta,\,(2)\eta'}(\omega) =
\tilde{A}^{(\mbsu{pp})}_{(11)\eta,\,(22)\eta'}(\omega)$,
\be
\tilde{A}^{\vphantom{(\mbsu{pp})}}_{(1)\eta,\,(2)\eta'}(\omega)
= - \frac{\eta\,\delta^{\vphantom{(J)}}_{\eta,\eta'}
\delta^{\vphantom{(J)}}_{(12)}}
{\omega - 2\,\eta\,E^{\vphantom{(J)}}_{(1)}}\,,
\label{aqrpa}
\ee
and ${\cal F}^{J=0\,(\mbsu{pp})}_{(11)\eta,\,(22)\eta'}$
is defined by Eq.~(C8) of Ref.~\cite{gmrQTBA}.

The matrix function
$\tilde{A}^{(\mbsu{pp})}_{(1)\eta,\,(2)\eta'}(\omega)$
has two poles at $\omega=0$ (first and second order ones)
and can be represented in the form:
\be
\tilde{A}^{(\mbsu{pp})}(\omega) =
\frac{a^{(2)}}{\omega^2} + \frac{a^{(1)}}{\omega} +
\tilde{A}^{\mbsu{(pp)\,reg}}(\omega)\,,
\label{apptr}
\ee
where function $\tilde{A}^{\mbsu{(pp)\,reg}}(\omega)$
is regular at $\omega \to 0$.
From the symmetry properties of the matrix function
$\tilde{A}^{(\mbsu{pp})}_{(1)\eta,\,(2)\eta'}(\omega)$
it follows that the matrices $a^{(2)}$ and $a^{(1)}$
are real, symmetric:
\be
a^{(2)} =\, a^{(2)*} =\, a^{(2)\,\mbsu{T}}\,,\qquad
a^{(1)} =\, a^{(1)*} =\, a^{(1)\,\mbsu{T}}\,,
\label{a21s}
\ee
and possess properties:
\be
a^{(2)} = \hat{\eta}^{\,x}a^{(2)}\hat{\eta}^{\,x}\,,\qquad
a^{(1)} = - \hat{\eta}^{\,x}a^{(1)}\hat{\eta}^{\,x}\,,
\label{a21sx}
\ee
where
\be
\hat{\eta}^{\,x}_{(1)\eta,\,(2)\eta'} =
\delta^{\vphantom{(J)}}_{\eta,-\eta'}\,
\delta^{\vphantom{(J)}}_{(12)}\,.
\label{etax}
\ee
Further, by substituting (\ref{apptr})
into Eq.~(\ref{appt1}) and putting the coefficients at the
same powers of $\omega$ to be equal to each other
we obtain the following equations for the matrices
$a^{(2)}$ and $a^{(1)}$:
\be
a^{(2)}
= - \tilde{A}(0)\,{\cal F}^{(\mbsu{pp})}\,a^{(2)}
= - a^{(2)}\,{\cal F}^{(\mbsu{pp})}\,\tilde{A}(0)\,,
\label{a2eq}
\ee
\be
\bigl(\,1 + \tilde{A}(0)\,{\cal F}^{(\mbsu{pp})}\bigr)\,a^{(1)}
= \tilde{A}(0)\,\hat{\eta}^{\,z}a^{(2)}\,,
\label{a1eq}
\ee
\be
a^{(2)} + a^{(2)}\hat{\eta}^{\,z}a^{(1)} = 0\,,
\label{a21eq}
\ee
where
\be
\hat{\eta}^{\,z}_{(1)\eta,\,(2)\eta'} =
\eta\,\delta^{\vphantom{(J)}}_{\eta,\eta'}\,
\delta^{\vphantom{(J)}}_{(12)}\,.
\label{etaz}
\ee

Eqs. (\ref{appt1}) and (\ref{apptr}) uniquely determine
the matrices $a^{(2)}$ and $a^{(1)}$.
However, strictly speaking, from this it does not follow that
a solution of Eqs. (\ref{a2eq})--(\ref{a21eq}) constrained by the
conditions (\ref{a21s}) and (\ref{a21sx}) is unique.
Nevertheless, we will find a particular solution of these equations
assuming that it is a true answer.
Calculations confirm the correctness of this assumption.

As follows from the above analysis
(see Eqs. (\ref{rgh1}) and (\ref{rgh2})),
the solution of Eqs.~(\ref{a2eq}) can be represented in the form:
\be
a^{(2)}_{(1)\eta,\,(2)\eta'} = a^{\vphantom{(J)}}_0\,
\rho^{\vphantom{(+)}}_{(1)\eta}\,
\rho^{\vphantom{(+)}}_{(2)\eta'}\,,
\label{a2sol1}
\ee
where $a^{\vphantom{(J)}}_0$ is a constant and
$\rho^{\vphantom{(+)}}_{(1)\eta}$ is defined by Eqs.~(\ref{rgh2}).
In the symbolic notations we have
\be
a^{(2)} = a^{\vphantom{(J)}}_0\,
|\rho\rangle\langle \rho |\,.
\label{a2sol2}
\ee
Let us search for the matrix $a^{(1)}$ in the form:
\be
a^{(1)} = \frac{a^{\vphantom{(J)}}_0}{2}\,
\bigl(\,|\rho\rangle\langle \xi | +
|\xi\rangle\langle \rho |\,\bigr)\,,
\label{a1sol2}
\ee
or in the explicit form:
\be
a^{(1)}_{(1)\eta,\,(2)\eta'} = \frac{a^{\vphantom{(J)}}_0}{2}\,
\bigl(\,
\rho^{\vphantom{(+)}}_{(1)\eta}\,
\xi^{\vphantom{(+)}}_{(2)\eta'} +
\xi^{\vphantom{(+)}}_{(1)\eta}\,
\rho^{\vphantom{(+)}}_{(2)\eta'}
\,\bigr)\,.
\label{a1sol1}
\ee
In addition, in agreement with Eqs. (\ref{rgh2}), (\ref{a21sx}),
(\ref{etax}), and (\ref{a1sol1}), we will assume
that the quantity $\xi^{\vphantom{(+)}}_{(1)\eta}$ does not depend
on $\eta$, i.~e.:
\be
\xi^{\vphantom{(+)}}_{(1)\eta} = \xi^{\vphantom{(+)}}_{(1)}\,.
\label{xieta}
\ee

By substituting (\ref{a2sol2}) and (\ref{a1sol2}) into
Eq.~(\ref{a1eq}) we obtain:
\be
\xi^{\vphantom{(+)}}_{(1)} = \sum_{(2)}
\bigl(\Omega^{-1}\bigr)_{(12)}\rho{\vphantom{\bigr)}}_{(2)}\,,
\label{xisol}
\ee
where
\be
\Omega{\vphantom{\bigr)}}_{(12)} =
E^{\vphantom{(J)}}_{(1)}\,\delta^{\vphantom{(J)}}_{(12)}
+ \frac{\sqrt{(2j^{\vphantom{(J)}}_1+1)(2j^{\vphantom{(J)}}_2+1)}}{8\pi}
\bigl(u^2_{(1)}-v^2_{(1)}\bigr)\bigl(u^2_{(2)}-v^2_{(2)}\bigr)
{\cal F}^{\xi}_{(11,22)}\,.
\label{omg}
\ee
Here
\be
{\cal F}^{\xi}_{(12,34)} =
\delta_{\tau_{\mbts{1}},\,\tau_{\mbts{2}}}
\delta_{\tau_{\mbts{3}},\,\tau_{\mbts{4}}}
\delta_{\tau_{\mbts{1}},\,\tau_{\mbts{3}}}
\int_0^{\infty}dr\,r^2
R^{\vphantom{(J)}}_{(1)}(r)R^{\vphantom{(J)}}_{(2)}(r)
R^{\vphantom{(J)}}_{(3)}(r)R^{\vphantom{(J)}}_{(4)}(r)
{\cal F}^{\xi}(r)\,,
\label{fxi}
\ee
$R^{\vphantom{(J)}}_{(1)}(r)$ is the radial part of the single-particle
wave function, the quantity ${\cal F}^{\xi}(r)$ determines the
effective residual interaction in the pp channel (see \cite{QTBA2,gmrQTBA}).

Substitution of Eqs. (\ref{a2sol2}) and (\ref{a1sol2}) into
Eq.~(\ref{a21eq}) yields:
\be
a^{-1}_0 = - \sum_{(1)}
\rho^{\vphantom{(+)}}_{(1)}\xi^{\vphantom{(+)}}_{(1)}\,.
\label{a0f}
\ee
Thus, the matrices $a^{(2)}$ and $a^{(1)}$ are completely
determined.

Let us introduce the following matrices:
\be
P = 1 + a^{(1)}\hat{\eta}^{\,z}\,,\qquad
P^{\dag} = 1 + \hat{\eta}^{\,z}a^{(1)}\,.
\label{defpp}
\ee
In the explicit form we have:
\bea
P^{\vphantom{\dag}}_{(1)\eta,\,(2)\eta'} &=&
\delta^{\vphantom{(J)}}_{\eta,\eta'}\,
\delta^{\vphantom{(J)}}_{(12)} +
\frac{a^{\vphantom{(J)}}_0}{2}\,
\bigl(\,\eta\,
\rho^{\vphantom{(+)}}_{(1)}
\xi^{\vphantom{(+)}}_{(2)}\eta' +
\xi^{\vphantom{(+)}}_{(1)}
\rho^{\vphantom{(+)}}_{(2)}
\,\bigr)\,,
\label{defp1d}\\
P^{\dag}_{(1)\eta,\,(2)\eta'} &=&
\delta^{\vphantom{(J)}}_{\eta,\eta'}\,
\delta^{\vphantom{(J)}}_{(12)} +
\frac{a^{\vphantom{(J)}}_0}{2}\,
\bigl(\,
\rho^{\vphantom{(+)}}_{(1)}\,
\xi^{\vphantom{(+)}}_{(2)} +
\eta\,\xi^{\vphantom{(+)}}_{(1)}\,
\rho^{\vphantom{(+)}}_{(2)}\eta'
\,\bigr)\,.
\label{defp2d}
\eea
In the general case of non-diagonal matrices these formulas
acquire the form:
\bea
P^{\vphantom{\dag}}_{(12)\eta,\,(34)\eta'} &=&
\delta^{\vphantom{(J)}}_{\eta,\eta'}\,
\delta^{\vphantom{(J)}}_{(13)}\,\delta^{\vphantom{(J)}}_{(24)} +
\delta^{\vphantom{(J)}}_{(12)}\,\delta^{\vphantom{(J)}}_{(34)}\,
a^{(1)}_{(1)\eta,\,(3)\eta'}\,\eta'
\label{defp1n}\\
P^{\dag}_{(12)\eta,\,(34)\eta'} &=&
\delta^{\vphantom{(J)}}_{\eta,\eta'}\,
\delta^{\vphantom{(J)}}_{(13)}\,\delta^{\vphantom{(J)}}_{(24)} +
\delta^{\vphantom{(J)}}_{(12)}\,\delta^{\vphantom{(J)}}_{(34)}\,
\eta\,a^{(1)}_{(1)\eta,\,(3)\eta'}\,.
\label{defp2n}
\eea

It is easy to show that the following equalities are fulfilled:
\be
PP=P\,,\qquad P^{\dag}P^{\dag}=P^{\dag}\,.
\label{ppr1}
\ee
Consequently, $P$ and $P^{\dag}$ are the projection operators.
In addition, we have:
\be
P\,a^{(1,2)} = \,0 = a^{(1,2)}P^{\dag}\,.
\label{ppr2}
\ee

Eqs. (\ref{apptr}) and (\ref{ppr2}) mean that one can use
the operators $P$ and $P^{\dag}$ to eliminate the coupling
of complex (2q$\otimes$phonon) configurations to the ghost state.
From the above analysis it follows that the elimination can be
achieved with the help of the replacement of the amplitude
$\bar{\Phi}(\omega)$ in formula (\ref{app2}) by the
projected amplitude $\bar{\Phi}^{\,\mbsu{proj}}(\omega)$
defined as
\be
\bar{\Phi}^{\,\mbsu{proj}}(\omega) =
P^{\dag}\,\bar{\Phi}(\omega)\,P\,.
\label{phiprj}
\ee
In what follows this method will be referred to as the QTBA
with $0^+$ projection.

\section{Calculations of the {\boldmath $\,0^+$} excitations
in the $\mathbf{^{120}Sn}$ nucleus}

As an illustration of the method described above consider results of
calculations of the $0^+$ excitations in the semi-magic nucleus
$^{120}$Sn. Calculational scheme is described in detail
in Ref.~\cite{gmrQTBA}.
It is based on the Hartree-Fock and BCS approximations
and is fully self-consistent on the RPA level.
The self-consistent mean field and the effective residual interaction
(including the spin-orbital and the Coulomb contributions
in both quantities) were derived from the Skyrme energy functional.
In the present calculations, the T5 Skyrme force parametrization
\cite{TBFP} was used. The single-particle continuum was included
completely on the RPA level.
The set of the phonons entering 2q$\otimes$phonon configurations
in the QTBA calculations included 29 collective modes
with values of the spin $L$ in the interval $2\leqslant L \leqslant 9$
and with natural parity $\pi = (-1)^L$.
The smearing parameter $\Delta$ in Eq.~(\ref{defsf}) is equal to
200~keV in all the calculations.


\begin{figure}[hb]
\includegraphics*[scale=0.7,angle=90]{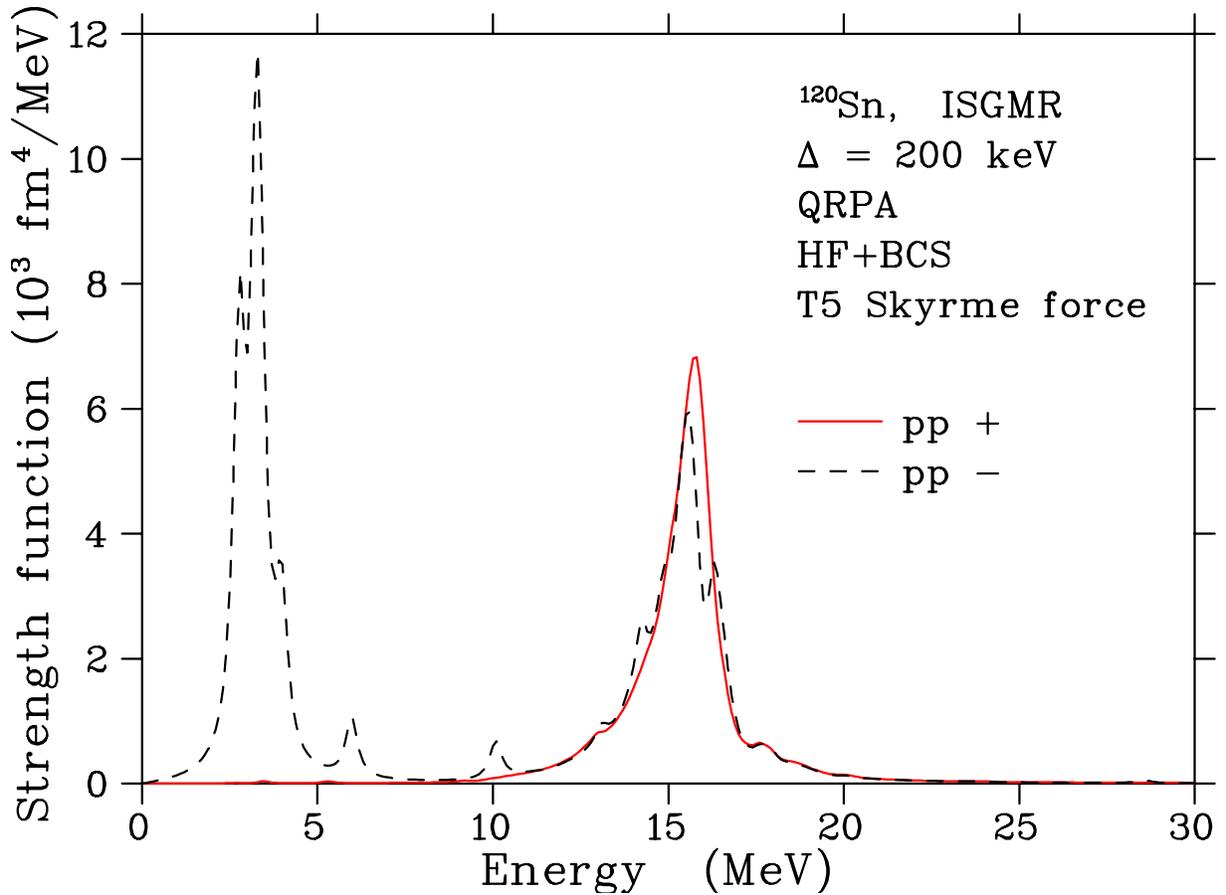}
\caption{\label{fig1}
Strength function of the
isoscalar giant monopole resonance in $^{120}$Sn
calculated within the QRPA with (solid line) and
without (dashed line) contributions of the pp channel.}
\end{figure}


\begin{figure}[!hb]
\includegraphics*[scale=0.7,angle=90]{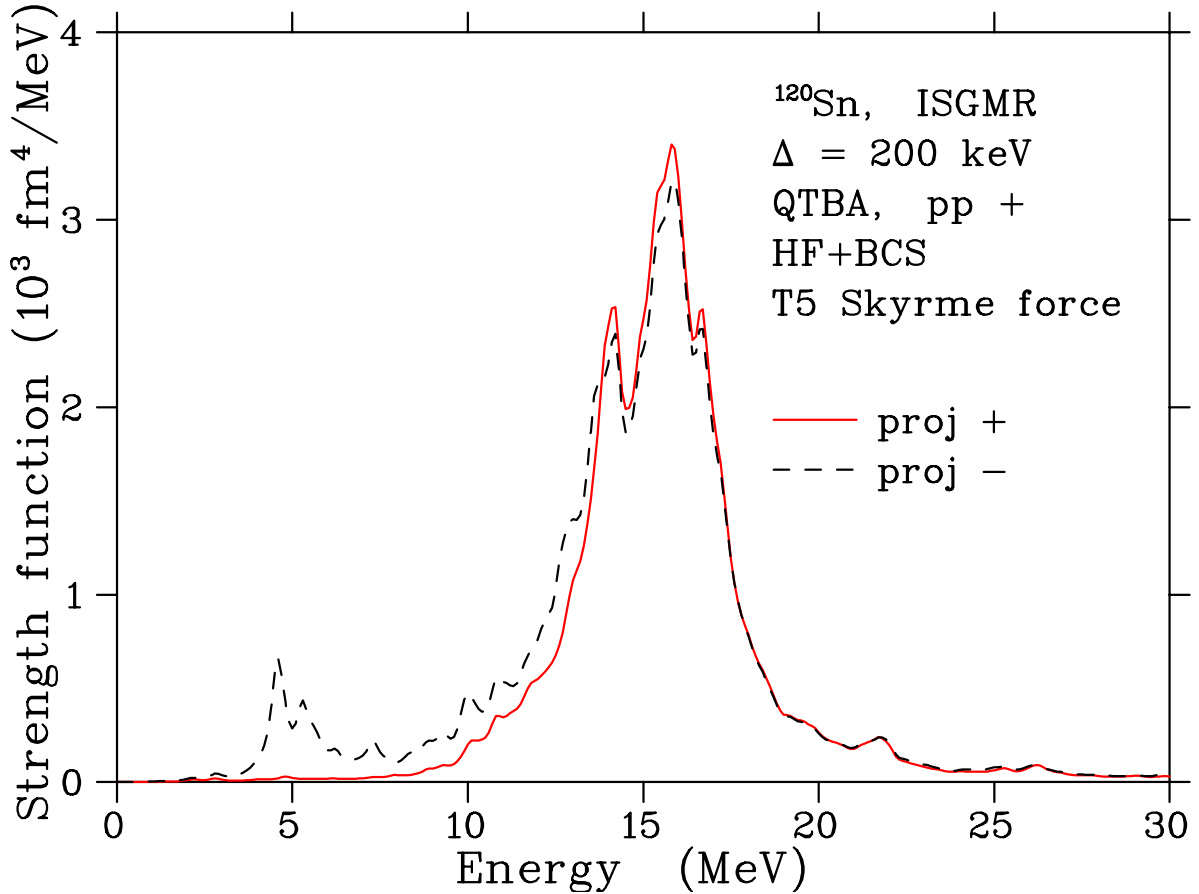}
\caption{\label{fig2}
Strength function of the
isoscalar giant monopole resonance in $^{120}$Sn
calculated within the QTBA including contributions of
the pp channel. The results with and without $0^+$ projection
are represented by the solid line and the dashed line, respectively.}
\end{figure}


\begin{figure}[!hb]
\includegraphics*[scale=0.7,angle=90]{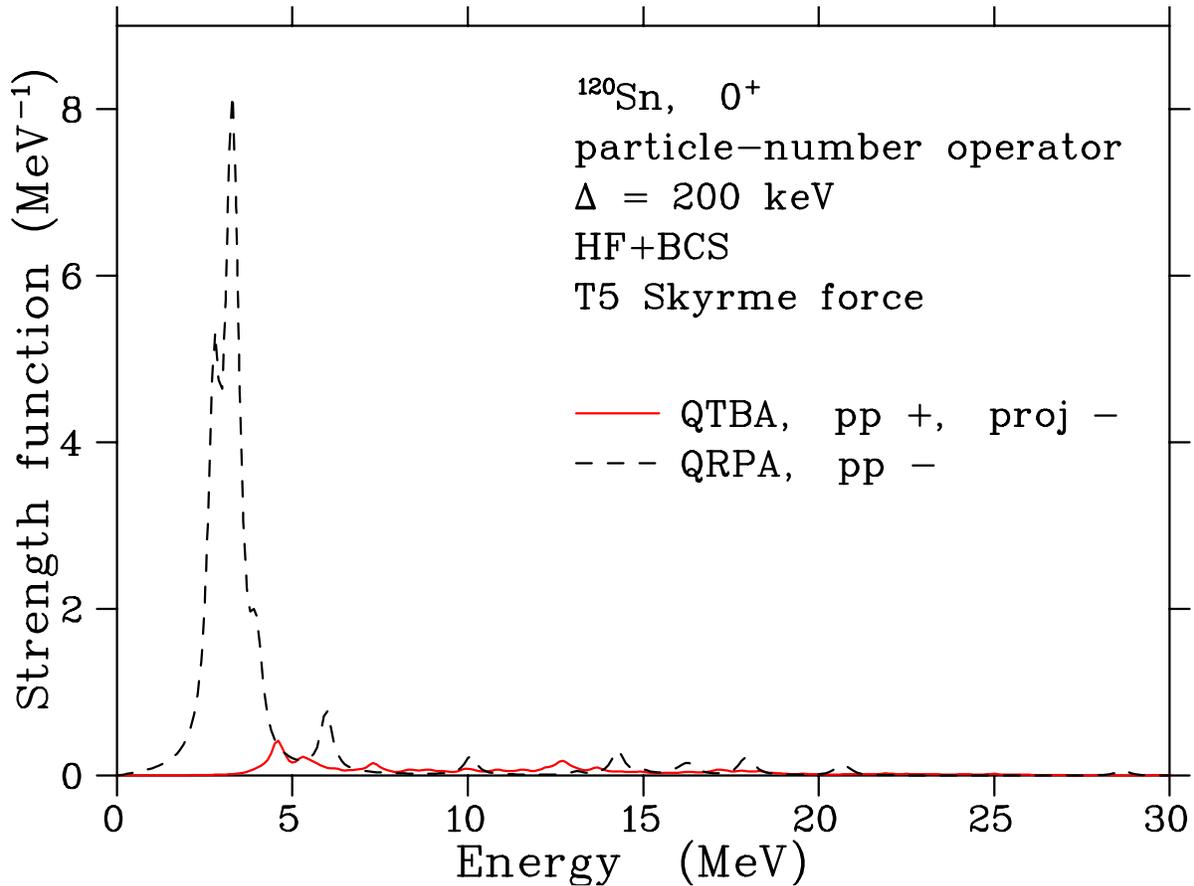}
\caption{\label{fig3}
Strength function of the spurious excitations in $^{120}$Sn
caused by the particle-number operator.
The solid line represents calculation within the QTBA including
contributions of the pp channel, but without $0^+$ projection.
The dashed line represents calculation within the QRPA
without contributions of the pp channel.}
\end{figure}

According to this scheme, the following sorts of calculations
were performed: (i) the QRPA and the QTBA calculations
with and without taking into account contributions of the pp channel
(they will be referred to as pp$+$ and pp$-$ calculations, respectively);
(ii) the QTBA calculations with and without $0^+$ projection
(they will be referred to as proj$+$ and proj$-$ calculations,
respectively).
In Figs. \ref{fig1} and \ref{fig2}, the results of calculations of
the isoscalar giant monopole resonance are shown.
In Fig.~\ref{fig1}, the results obtained within the QRPA (pp$+$)
and the QRPA (pp$-$) are presented.
Spurious $0^+$ excitations arise at low energies in the response
calculated without contributions of the pp channel.
These spurious excitations disappear when the pp-channel
contributions are included. The same picture is obtained in the
QTBA calculations.
However, to eliminate spurious excitations in the QTBA it is insufficient
to include the pp channel. Only the QTBA (pp$+$, proj$+$)
that includes pp-channel contributions together with $0^+$ projection
gives the $0^+$ response free from the spurious states.
In the calculations within QTBA (pp$+$, proj$-$) the fragmented
ghost states remain in the low-energy region, though their strength is
strongly suppressed as compared with calculations in which the pp-channel
contributions are not included (see Fig.~\ref{fig2}).

In Fig.~\ref{fig3}, the $0^+$ response to the particle-number operator
in $^{120}$Sn is shown. These results demonstrate that
the QRPA (pp$-$) and the QTBA (pp$+$, proj$-$) produce non-zero response.
In the calculations within the QRPA (pp$+$) and the QTBA (pp$+$, proj$+$)
this response disappears within calculational accuracy
as it should be.

\section{Conclusions}

In this work, the projection method is formulated which solves
the problem of the $0^+$ spurious states in the model intended
for the description of excitations in open-shell nuclei
with taking into account the single-particle continuum,
the pairing correlations, and the quasiparticle-phonon coupling.
The efficiency of the method is illustrated by the calculations
of the $0^+$ excitations in the $^{120}$Sn nucleus within
the self-consistent scheme based on the
Skyrme-Hartree-Fock approximation.

\begin{acknowledgements}
The work was supported by the Deutsche Forschungsgemeinschaft
under Grant No.~436~RUS~113/994/0-1 and by the
Russian Federal Agency of Education under project No.~2.1.1/4779.
\end{acknowledgements}


\begin{thebibliography}{99}
\bibitem{RS80}
P. Ring and P. Schuck,
{\it The Nuclear Many-Body Problem} (Springer-Verlag, New York, 1980).
\bibitem{T61}
D. J. Thouless,
Nucl. Phys. {\bf 22}, 78 (1961).
\bibitem{KS82}
V. A. Khodel and E. E. Saperstein,
Phys. Rep. {\bf 92}, 183 (1982).
\bibitem{QTBA1}
V. I. Tselyaev,
Phys. Rev. C {\bf 75}, 024306 (2007).
\bibitem{QTBA2}
E. V. Litvinova and V. I. Tselyaev,
Phys. Rev. C {\bf 75}, 054318 (2007).
\bibitem{M67}
A. B. Migdal,
{\it Theory of Finite Fermi Systems
and Applications to Atomic Nuclei}
(Interscience, New York, 1967).
\bibitem{SB75}
S. Shlomo and G. Bertsch,
Nucl. Phys. {\bf A243}, 507 (1975).
\bibitem{gmrQTBA}
V.~Tselyaev, J.~Speth, S.~Krewald, E.~Litvinova, S.~Kamerdzhiev,
N.~Lyutorovich, A.~Avdeenkov, and F.~Gr\"ummer,
Phys. Rev. C {\bf 79}, 034309 (2009).
\bibitem{PS88}
A. P. Platonov and E. E. Saperstein,
Nucl. Phys. {\bf A486}, 63 (1988).
\bibitem{TBFP}
F. Tondeur, M. Brack, M. Farine, and J. M. Pearson,
Nucl. Phys. {\bf A420}, 297 (1984).
%
\end{thebibliography}
\end{document}